
\magnification=\magstep1
\overfullrule=0pt
\vsize=46pc
\def\BZT{{\rm Z{\hbox to 3pt{\hss\rm Z}}}}
\def\BZS{{\hbox{\sevenrm Z{\hbox to 2.3pt{\hss\sevenrm Z}}}}}
\def\BZSS{{\hbox{\fiverm Z{\hbox to 1.8pt{\hss\fiverm Z}}}}}
\def\BZ{{\mathchoice{\BZT}{\BZT}{\BZS}{\BZSS}}}
\def\BCT{\,\hbox{\hbox to -3pt{\vrule height 6.5pt width
     .2pt\hss}\rm C}}
\def\BCS{\,\hbox{\hbox to -2.2pt{\vrule height 4.5pt width .2pt\hss}$
   \scriptstyle\rm C$}}
\def\BCSS{\,\hbox{\hbox to -2pt{\vrule height 3.3pt width
   .2pt\hss}$\scriptscriptstyle \rm C$}}
\def\BC{{\mathchoice{\BCT}{\BCT}{\BCS}{\BCSS}}}
\def\BRT{{\rm I{\hbox to 5.5pt{\hss\rm R}}}}
\def\BRS{{\hbox{\sevenrm I{\hbox to 4.3pt{\hss\sevenrm R}}}}}
\def\BRSS{{\hbox{\fiverm I{\hbox to 3.35pt{\hss\fiverm R}}}}}
\def\BR{{\mathchoice{\BRT}{\BRT}{\BRS}{\BRSS}}}
\def\BNT{{\rm I{\hbox to 5.5pt{\hss\rm N}}}}
\def\BNS{{\hbox{\sevenrm I{\hbox to 4.3pt{\hss\sevenrm N}}}}}
\def\BNSS{{\hbox{\fiverm I{\hbox to 3.35pt{\hss\fiverm N}}}}}
\def\BN{{\mathchoice{\BNT}{\BNT}{\BNS}{\BNSS}}}
\def\lb{\lbrack}
\def\rb{\rbrack}
\def\w{{\cal W}}
\def\section#1{\vbox{\leftline{\bf #1}\vskip-10pt\line{\hrulefill}}}
\def\subs#1{\vbox{\rightline{\bf #1}\vskip-10pt\line{\hrulefill}}}
\def\q#1{\lb#1\rb}
\def\bibitem#1{\parindent=8mm\item{\hbox to 6 mm{$\q{#1}$\hfill}}}
\def\mn{\medskip\smallskip\noindent}
\def\sn{\smallskip\noindent}
\def\bn{\bigskip\noindent}
\def\slu{$\widehat{sl(2,\BR)}_k/\widehat{U(1)}$}
\def\wfuenf{$\w(2,3,4,5)$}      
\def\wsieben{$\w(2,3,4,5,6,7)$} 
\def\trunc{\triangleright}                     
\def\truncrl{\hbox{$\,\triangleleft\triangleright\,$}} 
\def\sltr{\widehat{sl(2,\BR)}}
\def\slsl{$\widehat{sl(2,\BR)}_k\oplus
   \widehat{sl(2,\BR)}_{-{1\over2}}/\widehat{sl(2,\BR)}_{k-{1\over2}}$}
\def\orb#1{\hbox{Orb}\left(#1\right)}
\def\pl{\hbox{$+$}}
\def\Cdot{\hbox{$\cdot$}}
\def\Skp{\hskip 2pt}
\def\tru{1}
\def\winf{2}
\def\kd{3}
\def\bouschou{1}
\def\kauwatts{2}
\def\blm{3}
\def\rva{4}
\def\wirrep{5}
\def\howcl{6}
\def\kausch{7}
\def\flohr{8}
\def\bowwatts{9}
\def\FORT{10}
\def\fehort{11}
\def\dBT{12}
\def\bbss{13}
\def\ajl{14}
\def\hornfeck{15}
\def\klausREP{16}
\def\technical{17}
\def\bogo{18}
\def\altschuler{19}
\def\poly{20}
\def\bersh{21}
\def\altbausal{22}
\def\bouwknegt{23}
\def\andrdipl{24}
\def\popeo{25}
\def\popet{26}
\def\kac{27}
\def\matsuo{28}
\def\humph{29}
\def\fateev{30}
\def\wisski{31}
\font\large=cmbx10 scaled \magstep4
\font\largew=cmsy10 scaled \magstep4
\font\bigf=cmr10 scaled \magstep2
\pageno=0
\def\folio{
\ifnum\pageno<1 \footline{\hfil} \else\number\pageno \fi}
\phantom{not-so-FUNNY}
\rightline{ DFTT--15/94\break}
\rightline{ BONN--TH--94--01\break}
\rightline{ hep-th/9404113\break}
\rightline{ April 1994\break}
\vskip 1.0truecm
\centerline{\large Unifying {\largew W}-Algebras}
\vskip 1.0truecm
\centerline{\bigf R.\ Blumenhagen\raise5pt\hbox{\tenrm 1},
W.\ Eholzer\raise5pt\hbox{\tenrm 1},}
\medskip
\centerline{\bigf A.\ Honecker\raise5pt\hbox{\tenrm 1},
K.\ Hornfeck\raise5pt\hbox{\tenrm 2},
R.\ H\"ubel\raise5pt\hbox{\tenrm 1}
}
\bigskip\medskip
\centerline{\it ${}^{1}$ Physikalisches Institut der Universit\"at Bonn}
\centerline{\it Nu{\ss}allee 12, 53115 Bonn, Germany}
\medskip
\centerline{\it ${}^{2}$ INFN, Sezione di Torino}
\centerline{\it Via Pietro Giuria 1, 10125 Torino, Italy}
\vskip 1.1truecm
\centerline{\bf Abstract}
\vskip 0.2truecm
\noindent
We show that quantum Casimir $\w$-algebras truncate at degenerate
values of the central charge $c$ to a smaller algebra if the rank
is high enough: Choosing a suitable parametrization of the
central charge in terms of the rank of the underlying
simple Lie algebra,
the field content does not change with the rank of the Casimir
algebra any more. This leads to identifications between the
Casimir algebras themselves but also gives rise to new,
`unifying' $\w$-algebras. For example, the $k$th unitary
minimal model of ${\cal WA}_n$ has a unifying $\w$-algebra
of type $\w(2,3,\ldots,k^2 + 3 k + 1)$.
These unifying $\w$-algebras are
non-freely generated on the quantum level and belong to a
recently discovered class of $\w$-algebras with infinitely,
non-freely generated classical counterparts. Some of the
identifications are indicated by level-rank-duality leading
to a coset realization
of these unifying $\w$-algebras. Other unifying $\w$-algebras
are new, including e.g.\ algebras of type ${\cal WD}_{-n}$.
We point out that all unifying quantum $\w$-algebras are
{\it finitely}, but non-freely generated.
\vfill
\eject
\section{0.\ Introduction}
\mn
Quantum $\w$-algebras play a fundamental r\^ole in
two dimensional conformal field theory (for a recent review see
e.g.\ $\q{\bouschou}$). In this letter we shall focus
on the so-called `generic' (or `deformable')
$\w$-algebras where the structure constants
are continuous functions (with isolated singularities) of the
Virasoro centre $c$ for a given set of generating fields. There
is also the class of `non-deformable' $\w$-algebras that exist
only for finitely many isolated values of the central charge $c$.
Non-deformable $\w$-algebras with few generators and their
representations have been extensively studied in $\q{\kauwatts-\howcl}$.
On the basis of these results it is now generally believed
that non-deformable $\w$-algebras can be
regarded as extensions or truncations of deformable ones
\footnote{${}^{1})$}{Note that in $\q{\howcl}$ there is a misleading remark
on this question because it was not recognized that
$\w(2,8)$ at $c=-{712 \over 7}$ and $c=-{3164 \over 23}$
arises as a truncation of ${\cal WE}_8$ and ${\cal WE}_7$
respectively.} -- with known exceptions $\q{\kausch,\flohr}$.
\mn
The finitely generated generic quantum $\w$-algebras fall at least
into two classes. The class that consists of freely generated
$\w$-algebras is already quite well understood. There
are indications $\q{\bowwatts-\fehort}$ that all algebras in this
class can be obtained by quantized $\q{\dBT}$ Drinfeld-Sokolov
reduction (see e.g.\ $\q{\FORT}$ and references therein). In
particular, quantized Drinfeld-Sokolov reduction
for a principal $sl(2)$ embedding into a simple
Lie algebra ${\cal L}_n$ of rank $n$ leads to the
so-called `Casimir algebras' $\q{\bbss}$ which we denote by
`${\cal WL}_n$' (we will also call $n$ the `rank of ${\cal WL}_n$').
Note that the classical counterparts of the $\w$-algebras in this
class are {\it finitely, freely} generated.
\mn
It was recently shown in $\q{\ajl}$ that there is another class of
$\w$-algebras with {\it infinitely, non-freely} generated classical
counterparts. Due to cancellations between generators and relations
upon normal ordering, the quantized versions of these $\w$-algebras
become finitely, non-freely generated in all known examples. This class
of $\w$-algebras contains --among others-- orbifolds and cosets
\footnote{${}^{2})$}{On the classical level all cosets belong to
this class. However, particular quantum coset constructions
(e.g.\ of Casimir algebras) yield finitely, {\it freely} generated
$\w$-algebras.}.
Two unexpected examples of finitely, but non-freely generated quantum
$\w$-algebras
had been found earlier: Particular extensions of the Virasoro algebra
with additional fields of conformal dimensions 4, 6 $\q{\kauwatts}$
and 3, 4, 5 $\q{\hornfeck}$, respectively. In particular, the origin
of the former one (which we denote by `$\w(2,4,6)$') was mysterious
for some time $\q{\howcl}$ although it was proposed in $\q{\klausREP}$
to identify this algebra formally with `${\cal WD}_{-1}$'.
In $\q{\ajl,\technical}$ these $\w(2,4,6)$ and $\w(2,3,4,5)$ were
explained in terms of the cosets \slsl\ and \slu.
\mn
In this letter we will show that this second class contains
`unifying $\w$-algebras' which interpolate
the rank $n$ of Casimir algebras ${\cal WL}_n$ at particular values
of the central charge. For example,
the $\w(2,3,4,5)$ unifies the first unitary minimal
models of ${\cal WA}_{n-1}$ whereas the $\w(2,4,6)$ corresponds
to certain minimal models of ${\cal WC}_n$. Using the coset
realization of these algebras, the unifying $\w$-algebras can
be regarded as a generalization of level-rank-duality
$\q{\bogo,\altschuler}$. These identifications between a priori
different $\w$-algebras for particular values of the central
charge $c$ are closely related to the fact that for certain values of
the central charge $c$ some generators become null fields leading
to a `truncation' of the $\w$-algebra.
We will further show from inspection
of the Kac determinant for Casimir $\w$-algebras that
truncations of them are in fact a very general phenomenon.
These truncations of Casimir
$\w$-algebras imply truncations of the various linear $\w_\infty$-algebras
to a finitely generated algebra which can indeed be verified.
\mn
\subs{Unifying $\w$-algebras}
\sn
We introduce the notion of `unifying $\w$-algebras' before
establishing their existence. A unifying $\w$-algebra is a finitely generated
deformable quantum $\w$-algebra (i.e.\ existing for generic $c$) with the
following properties:
\sn
\item{$\bullet$}
There exists an integer $n_0$ and simple Lie algebras ${\cal L}_n$
such that for all $n \ge n_0$ and
$c=c_{{\cal L}_n}(p(n),q(n))$ it is isomorphic to the Casimir
algebra ${\cal WL}_n$.
\item{$\bullet$}
${\cal L}_n$ is either of type ${\cal A}$, ${\cal B}$, ${\cal C}$
or ${\cal D}$ independent of $n$.
\sn
This implies that
for all $n > n_0$ the Casimir algebras ${\cal WL}_n$ at central charge
$c_{{\cal L}_n}(p(n),q(n))$ truncate to the unifying $\w$-algebra
whose spin content is independent of $n$.
\sn
Since unifying $\w$-algebras exist for generic $c$, they
can be considered as continuations to real values of the
rank $n$ of Casimir algebras ${\cal WL}_n$ at $c_{{\cal L}_n}(p(n),q(n))$.
Some Casimir algebras play the r\^ole of unifying $\w$-algebras. In contrast
to them, the `true' unifying $\w$-algebras are non-freely generated and
belong to the second class of quantum $\w$-algebras.
\bn
\section{\tru.\ Truncations and identifications from structure constants}
\mn
The first unitary minimal model of ${\cal WA}_{k-1}$ has the symmetry algebra
$\w(2,3,4,5) \cong \sltr_k / \widehat{U(1)}$ {\it independent} of $k$
$\q{\ajl,\technical}$. This means that in the algebra
${\cal WA}_{k-1}$ all simple fields of dimension $\geq 6$ become
null fields and the algebra ${\cal WA}_{k-1}$ truncates
at $c_{{\cal A}_{k-1}}(k+1,k+2)={2(k-1) \over k+2}$ to the
algebra $\w(2,3,4,5)$ which is a {\it unifying} algebra for the
first unitary model of all ${\cal WA}_{k-1}$. In the following
we will investigate more general cases such as the $n$th unitary
model of ${\cal WA}_{k-1}$. Similarly, the coset
$\w_3^{(2)} / \widehat{U(1)} \cong \w(2,3,4,5,6,7)$ which will
be discussed in $\q{\technical}$ (`$\w_3^{(2)}$' denotes the
algebra of Polyakov and Bershadsky $\q{\poly,\bersh}$) is a unifying algebra
for the non-unitary minimal models $c_{{\cal A}_{k-1}}(k+1,k+3)$ of
${\cal WA}_{k-1}$.
\mn
In order to study these truncations and identifications in a
systematic way we make use of the observation $\q{\klausREP}$
that deformable $\w$-algebras can be
parametrized with `universal' formulae in which only the dimension
of a certain representation enters as parameter.
More precisely, the structure constants of
many $\w$-algebras fall into different classes, so that in each class the
structure constants can be described by universal formulae
$C_{n\,m}^l(c,h(c))$. The parameter $h(c)$ varies for
each of these algebras, being in fact any one of the two dimensions
which correspond to representations with a particularly low-lying
null-state. Usually, there is also a third parametrization $h(c)$
that does not correspond to degenerate representations.
\sn
One of these classes are $\w$-algebras of type $\w(2,3,4,\ldots)$,
that is $\w$-algebras with primary fields of dimensions 3 and 4 and
any number of fields with dimension $\geq 5$. Therefore,
all ${\cal WA}_{n-1}$-algebras and all truncations of them including
in particular the special algebras $\w(2,3,4,5) \cong
\sltr_k / \widehat{U(1)}$ and $\w(2,3,4,5,6,7) \cong
\w_3^{(2)} / \widehat{U(1)}$ belong to this class.
Another class is formed of some $\w$-algebras of type $\w(2,4,\ldots)$
which includes the algebras ${\cal WB}_n$ and ${\cal WC}_n$ as well as
the orbifolds of ${\cal WD}_n$ and ${\cal WB}(0,n)$. In this approach,
the structure constants of the algebras ${\cal WD}_n$ can be continued
to negative values of the rank such that the algebra
$\w(2,4,6) \cong {\cal WD}_{-1}$ belongs to this class $\q{\klausREP}$.
In $\q{\technical}$ we shall give a realization of ${\cal WD}_{-n}$
in terms of the coset $\widehat{sp(2n)}_k \oplus\widehat{sp(2n)}_{-{1 \over2}}
/ \widehat{sp(2n)}_{k-{1 \over 2}}$ and show that they are of type
$\w(2,4,\ldots,2n(n+2))$.
\sn
The fact that the structure constants in each class
can be described solely by the set of values of the
three parameters for $h(c)$, $H = \{h_1(c), h_2(c), h_3(c)\}$, has
the immediate consequence that the structure constants for two algebras
$X$ and $V$, and hence the algebras, become identical,
whenever the sets $H_{X}$ and $H_{V}$ are equal.
In general, the identification will lead
to a truncation of the algebra with the higher number of
simple fields, say $X$, to an algebra $V$ with
less fields at a {\it finite} set of values for
the central charge. We shall denote this truncation by the
symbol `$\trunc$', i.e.\
$$X\,\trunc\,V \quad{\rm if}\ \ H_{X} = H_{V}.
  \eqno{\rm(\tru.1)}$$
However, we remark that the two sets $H_X$ und $H_V$ are also equal if $V$
is a subalgebra of $X$ (at some fixed $c$-values) and no truncation takes
place. There are indeed special cases known where this occurs.
Thus, one has to prove the existence of null fields in order to confirm the
truncation expected from this argument. Indications for the presence
of null fields will be presented in later sections.
\sn
Note that (\tru.1) is usually automatically satisfied if one
$h_i \in H_{X}$ is equal to one $h_j \in H_{V}$. The exceptions
are related to singularities that occur e.g.\ at $c=0$ or at
$c=-2$ (see (\tru.2)).
\mn
\subs{Identifications of $\w$-algebras of type $\w(2,3,4,\ldots)$}
\sn
The first few structure constants for these algebras are given in
$\q{\klausREP}$. It is immediately clear from the structure constants
that any $\w$-algebra of type $\w(2,3,4,\ldots)$
(which can be described by these structure constants)
truncates to the algebra $\w(2,3)$ at $c=-2$:
$$\left.\matrix{
{\cal WA}_{n-1} & \trunc & {\cal W}(2,3) \cr
{\cal W}(2,3,4,5) & \trunc & {\cal W}(2,3) \cr
{\cal W}(2,3,4,5,6,7) & \trunc & {\cal W}(2,3) \cr}\right\}
\hskip0.4cm \hbox{at}\ \ c = -2 .
  \eqno{\rm (\tru.2)}$$
These truncations are independent of
the values of $h$ whence in these cases
it is not necessary that the corresponding sets $H$ agree.
We will omit such cases in the following and restrict our
discussion to cases where the sets $H$ agree.
\sn
For the three algebras mentioned the set
$H=\{h_1,h_2,h_3\}$ consists of the following values:
$$\eqalign{
{\cal WA}_{n-1}:&\quad
 4 n^2 h_{1,2}^2 + 2
\bigl( c\!-\!(n\!-\!1)(2n\!+\!1)\bigr)h_{1,2}\!+\!c(n\!-\!1)=0;\quad
h_3 ={c(n+1)\over{2(c\!+\!1\!-\!n)}} \cr
\w(2,3,4,5):&\quad h_1 = {3c\over{2(c+1)}};\quad
h_2={{c(2-c)}\over{8(c+1)}}; \qquad \qquad \,
h_3 = -{{c+4}\over{2(c+1)}} \cr
\w(2,3,4,5,6,7):&\quad h_1 ={{3(k+1)}\over{2k+3}};\quad
h_2= {{3(k+1)(k+2)}\over{2(2k+3)}};\quad
h_3=-{{(k+1)^2}\over{(k+3)(2k+3)}} \cr
&\quad \hbox{with}\ \ c =  -6{{(k+1)^2}\over{k+3}}. \cr
}\eqno{\rm (\tru.3)}$$
By identifying the sets $H_{{\cal WA}_{m-1}}$
and $H_{{\cal WA}_{n-1}}$ we find a solution for
the central charge where ${\cal WA}_{m-1}$ truncates
to ${\cal WA}_{n-1}$ ($m>n$):
$${\cal WA}_{m-1} \trunc {\cal WA}_{n-1} \hskip1cm \hbox{at}\ \
c_m(n)=  - {{(m-1)(n-1)(m+n+mn)}\over{m+n}} .
  \eqno{\rm(\tru.4)}$$
The identification (\tru.4) has already been proposed
in $\q{\altbausal}$ where equality of characters was
shown using the coset realization of ${\cal WA}_{n-1}$.
Let us now investigate the truncations of ${\cal WA}_{n-1}$
to \wfuenf. From eq.\ (\tru.3) we find the solutions
\footnote{${}^3)$}{For $n\leq 4$ \wfuenf\ truncates to
${\cal WA}_{n-1}$, for $n=5$ the two algebras are isomorphic
for the given values of $c$. We use therefore the symbol
`$\truncrl$'.}
$${\cal WA}_{n-1} \truncrl \, \w(2,3,4,5) \cong
  {\sltr_n \over \widehat{U(1)}}
\hskip1cm \hbox{at}\ \ \
\matrix{
c(n) = {{2(n-1)}\over{n+2}}\cr
c(n) = -1-3n\cr
c(n) = {{2(1-2n)}\over{n-2}} \, .
}\eqno{\rm(\tru.5)}$$
The first truncation is the well-known identification of the
coset $\sltr_n / \widehat{U(1)}$ with the ${\cal WA}_{n-1}$-algebra
for the first unitary minimal model.
The other two truncations do not correspond to minimal models.
\sn
We proceed in the same way for truncations
of ${\cal WA}_{n-1}$ to \wsieben :
$${\cal WA}_{n-1}\truncrl \, \w(2,3,4,5,6,7) \cong
   {{\cal W}_3^{(2)} \over \widehat{U(1)}}
\hskip1cm \hbox{at}\ \ \ \matrix{
k(n) = {{ -3(n+1)}\over{n+2}} &
c(n) = {{-2{{(2n+1)}^2}}\over{n+2}} \cr
k(n) = {{n-3}\over2 } &
c(n) = {{-3{{(n-1)}^2}}\over{n+3}} \cr
k(n) = {{3-2n}\over{n-2}} &
c(n) = {{6{{(n-1)}^2}}\over{(2-n)(n-3)}}\ .\cr}
\eqno{\rm (\tru.6)}$$
Here, we denote the level entering $\w_3^{(2)}$ by `$k$'
$\q{\technical,\poly,\bersh}$. Note that the second line
of (\tru.6) corresponds to minimal models of $\w_3^{(2)}$.
\mn
\subs{Identification of $\w$-algebras of type $\w(2,4,\ldots)$}
\sn
We will now look for identifications between algebras of type
$\w(2,4,\ldots)$. Since the bosonic projections
of the superalgebras ${\cal WB}(0,m)$  are
described by the formulae for the orbifold
of ${\cal WD}_n$ with $n=m+{1\over2}$ half-integer,
we shall not consider these bosonic projections separately, but
use the notation of the orbifold of ${\cal WD}_n$ for
both of them, having in mind that we treat the superalgebra
${\cal WB}(0,n-{1\over 2})$ whenever $n$ is half-integer.
A further argument for this notation is that both algebras are
realized in terms of diagonal $so(k)$-cosets with even and odd
$k=2n$ respectively (see e.g.\ $\q{\bouschou}$).
The orbifold of ${\cal WD}_{3\over 2}$ for example
is the bosonic projection of the $N=1$ Super Virasoro algebra --
a non-freely generated algebra of type $\w(2,4,6)$ (see e.g.\
$\q{\bouwknegt,\ajl}$).
Moreover, also negative values of $n$ are allowed for ${\cal WD}_{n}$.
\mn
The first structure constants for this class of $\w$-algebras have been
presented in $\q{\klausREP}$. We give only the classifying sets $H$:
$$\eqalign{
\orb{{\cal WD}_n}:\ \ &4n(n-1) h_{1,2}^2+
\bigl(c+n(3-4n)\bigr)h_{1,2}+c(n-1)=0; \quad  h_3 = n; \cr
{\cal WB}_n,{\cal WC}_n:\ \ & 8n(1+n) h_1^2+
                2\bigl(c-n(3+2n)\bigr)h_1+c(2n-1) = 0; \cr
 &(2n-1)(2n+1)h_2^2 + \bigl(c-n(6n+1)\bigr) h_2 +cn + 2n^2 = 0;\cr
 &2(c+2n) h_3^2 - \bigl(n(2n+3)+c(4n+3)\bigr) h_3+c(n+1)(2n+1) =  0; \cr
&\hbox{where one of the solutions is for $\cal WB$,
    the other one for $\cal WC$.}\cr
}\eqno{\rm(\tru.7)}$$
For this class of algebras we find the following
truncations with generic $m > n$:
$$\eqalignno{
\orb{{\cal WD}_m} &\trunc \orb{{\cal WD}_n} \hskip0.9cm \hbox{at}\ \ \
c=-{{mn(3-4m-4n+4mn)}\over{m+n-1}}, & {\rm(\tru.8)} \cr
{\cal WC}_m &\trunc {\cal WC}_n \hskip2cm \hbox{at}\ \ \
c={{mn(3+2m+2n-4mn)}\over{1+m+n}}, & {\rm(\tru.9)} \cr
{\cal WB}_m &\trunc {\cal WB}_n\hskip2cm \hbox{at}\ \ \
c=-{{(2m\!+\!n\!+\!2mn)(m\!+\!2n\!+\!2mn)}\over{m+n}}, & {\rm(\tru.10)} \cr
{\cal WB}_m &\trunc {\cal WC}_n\hskip2cm \hbox{at}\ \ \
c={{4mn(3+2m+2n)}\over{(1+2m-2n)(1-2m+2n)}}. & {\rm (\tru.11)} \cr}$$
Finally, for generic $m$ and $n$, one has the following
identifications where the larger algebra truncates to the smaller one:
$$\eqalignno{
{\cal WB}_m &\truncrl \, {\cal WC}_n \hskip1cm \hbox{at}\ \ \
c=-{{2mn(3+2m+2n+4mn)}\over{1+2m+2n}},& {\rm(\tru.12)} \cr
\orb{{\cal WD}_m} &\truncrl \, {\cal WB}_n\hskip1cm \hbox{at}\ \ \
c={{2mn(-3+4m-2n+4mn)}\over{1-2m-2n}},& {\rm(\tru.13)} \cr
\orb{{\cal WD}_m} &\truncrl \, {\cal WB}_n\hskip1cm \hbox{at}\ \ \
c={{(n\!-\!m\!-\!2mn)(2mn\!-\!m\!-\!2n)}\over{m+n}}, & {\rm(\tru.14)} \cr
\orb{{\cal WD}_m} &\truncrl \, {\cal WC}_n\hskip1cm \hbox{at}\ \ \
c={{mn(3-4m+2n)}\over{(1-m+n)(1-2m+2n)}}. & {\rm(\tru.15)} \cr
}$$
For special values of $m$ and $n$ one might find additional solutions.
\mn
Let us now discuss some interesting consequences of these
identifications. Eq.\ (\tru.14) can be parametrized by the formula
for the minimal models of ${\cal WB}_n$ as
$c_{{\cal B}_n}(2n,2n+2(m-{1\over 2})+1)$.
Thus, for half-integer $m=k+{1\over 2}$ ${\cal WB}_n$ truncates
at $c_{{\cal B}_n}(2n,2n+2k+1)$ to $\orb{{\cal WD}_m}=
\orb{{\cal WB}(0,k)}$. We can interpret $\orb{{\cal WB}(0,k)}$
as the unifying algebra of the ${\cal WB}_n$ models at
$c_{{\cal B}_n}(2n,2n+2k+1)$. Eq.\ (\tru.13)
can be parametrized as $c_{{\cal B}_n}(2n-1+2m,2n+1)$.
We infer that ${\cal WB}_n$ truncates at these values to
$\orb{{\cal WD}_m}$ which serves as a unifying
algebra for these ${\cal WB}_n$ minimal models.
Using the realization of ${\cal WD}_m$ and $\orb{{\cal WB}(0,m)}$
in terms of $so(k)$-cosets these identifications can be
summarized as follows:
$$\eqalign{
{\cal WB}_n \cong &
\cases{
{\widehat{so(k)}_\mu \oplus \widehat{so(k)}_1 \over
\widehat{so(k)}_{\mu+1} } \cong \orb{{\cal WB}(0,{k-1 \over 2})},
  & $k$ odd \cr
\orb{{\widehat{so(k)}_\mu \oplus \widehat{so(k)}_1 \over
\widehat{so(k)}_{\mu+1} }} \cong \orb{{\cal WD}_{k \over2}},
  & $k$ even \cr
} \cr
&\qquad\qquad \hbox{at} \quad \cases{
c={k n (3 + 2 n - 2 k - 2 k n) \over k + 2 n - 1},
&$\mu = -{2 n (k-2) \over 2 n + 1}$, \cr
c=-{(2 k n + k - 2 n) (2 k n - k - 4 n) \over 2 (k + 2 n)},
&$\mu = -{(2 n - 1) k - 4 n \over 2 n}$. \cr
}
}   \eqno({\rm \tru.16})$$
Note also that when replacing $m$ by $-m$, eq.\ (\tru.15) is parametrized
by $c_{{\cal C}_n}(m+n+1,2m+2n+1)$ which can be interpreted
as a truncation to the algebra $\orb{{\cal WD}_{-m}}$ $\q{\technical}$.
\mn
Interesting consequences follow from the fact that also the algebra
$\w(2,4,6,8,10)$ which arises as the orbifold of the coset
\slu\ $\q{\technical}$ (hence the orbifold of
the first unitary model of ${\cal WA}_{k-1}$) belongs to the class
of $\w$-algebras of type $\w(2,4,\ldots)$ with
$$h_1 = {{2-c} \over {3}}; \quad
h_2 = {{c+1} \over {2 \, (2-c)}}; \quad
h_3 = {{3 \,c} \over {2 \, (c+1)}} .
         \eqno({\rm \tru.17})$$
Note that in contrast to this the orbifold of ${\cal WA}_{n-1}$ for $c$
generic does not belong to this class, i.e.\ cannot be described by the
general structure constants. Proceeding in the same way as for the
previous truncations we find
$$\eqalignno{
\orb{{\cal WD}_m}&\trunc
\w(2,4,6,8,10) \cong \orb{{\sltr_k \over \widehat{U(1)}}}
\hskip0.7cm \hbox{at}\ \ \
\matrix{c(m) = 2 - 3 m  \cr
        c(m) = {{4m-1} \over {2m+1}}  \cr
        c(m) = - {{2 m} \over {2m-3}},  \cr
}         &({\rm \tru.18}) \cr\cr
{\cal WB}_m &\trunc
\w(2,4,6,8,10) \cong \orb{{\sltr_k \over \widehat{U(1)}}}
\hskip0.7cm \hbox{at}\ \ \
\matrix{c(m) = {{4  m} \over {2  m +3}} \cr
        c(m) = -2  m \cr
        c(m) = -  {{m+2} \over {m-1}}, \cr
}         &({\rm \tru.19}) \cr\cr
{\cal WC}_m &\trunc
\w(2,4,6,8,10) \cong \orb{{\sltr_k \over \widehat{U(1)}}}
\hskip0.7cm \hbox{at}\ \ \
  c(m) = -1 -6  m .\quad
         &({\rm \tru.20}) \cr
}$$
These truncations are compatible with the truncation of ${\cal WA}_{n-1}$ to
$\w(2,3,4,5)$ (see eq.\ (\tru.5)) exactly at the $c$-values for the first
unitary model of ${\cal WA}_{n-1}$. In particular,
we find that the orbifold of the first unitary model of ${\cal WA}_{2n-1}$
is equal to the orbifold of the second unitary model
of ${\cal WD}_{n \over 2}$ (this can be confirmed using level-rank duality
$\q{\technical}$), and that the orbifold of the first unitary model of
${\cal WA}_{2n}$ is equivalent to a unitary model of ${\cal WB}_n$. For
the example $n=2$, $c={8 \over 7}$ a detailed verification of this
truncation $\orb{\w(2,3,4,5)} \trunc \w(2,4)$ is possible $\q{\technical}$.
Note that this model is probably the only unitary minimal model of
$\w(2,4) \cong {\cal WB}_2$ $\q{\andrdipl}$.
\bn
\section{\winf.\ Null fields in linear $\w_{\infty}$-algebras}
\mn
In the previous section we have found identifications between $\w$-algebras.
They imply truncations for which we will provide further support in this
section by verifying that in the limit $n \to \infty$ one does indeed obtain
truncations of the various linear $\w_{\infty}$-algebras. There are four
types of linear $\w_{\infty}$-algebras, the usual $\w_{\infty}$,
$\w_{1+\infty}$ $\q{\popeo,\popet}$ and their subalgebras ${\cal WB}_{\infty}$
and ${\cal WC}_{\infty}$ formed by the even-dimensional fields.
The results in the previous section predict the following truncations:
$$\matrix{
({\rm \tru.2}) \ \Rightarrow \quad &
{\w}_{\infty} &\trunc &{\w}(2,3) & \hbox{at}\ \ \ c=-2 &\cr
({\rm \tru.5}) \ \Rightarrow \quad &
{\w}_{\infty} &\trunc &\w(2,3,4,5) & \hbox{at}\ \ \ c=-4
&\hbox{and}\ \ \ c=2 \cr
({\rm \tru.6}) \ \Rightarrow \quad &
{\w}_{\infty} &\trunc &\w(2,3,4,5,6,7) & \hbox{at}\ \ \ c=-6 &\cr
({\rm \tru.11}) \ \Rightarrow \quad &
{\cal WB}_{\infty} &\trunc & {\cal WC}_n & \hbox{at}\!\ \  c=-2n &\cr
({\rm \tru.15}) \ \Rightarrow \quad &
{\cal WC}_{\infty} &\trunc &\orb{{\cal WD}_n} &
\hbox{at}\ \ \ \, \ c=n &\cr
({\rm \tru.19}) \ \Rightarrow \quad &
{\cal WB}_{\infty} &\trunc &\w(2,4,6,8,10) & \hbox{at}\ \ \ c=-1
&\hbox{and}\ \ \ c=2 . \cr }
\eqno{\rm(\winf.1)}$$
The OPEs for the $\w_{\infty}$-algebras are particularly simple and
one can calculate the first null fields explicitely. In our calculations
we have restricted to the first null field which indicates but does
not really prove the truncation of a $\w_{\infty}$-algebra to a
finitely generated $\w$-algebra.
\mn
For $\w_{\infty}$ and ${\cal WB}_{\infty}$ we
have checked the presence of null fields up to scale dimension 12, for
${\cal WC}_{\infty}$ up to dimension 18. One
does indeed find null fields for those values of $c$ and the scale
dimension predicted by (\winf.1).
In addition to the confirmation of (\winf.1) we find further truncations.
For $\w_{\infty}$ we find that $\w_{\infty} \trunc \w(2,3,\ldots,9)$
at $c=-8$ and $\w_{\infty} \trunc \w(2,3,\ldots,11)$ at
$c=-10$ and $c=4$. On the basis of these truncations we conjecture that
for any integer $r$ a generic algebra of type $\w(2,3,4,\ldots,2r+1)$
exists which is a unifying algebra for ${\cal WA}_{n-1}$ at
$c_{{\cal A}_{n-1}}(n-r,n-r+1)$ (the conjecture for the
central charge is based on the cases $1 \le r \le 3$).
The truncation $\w_{\infty} \trunc \w(2,3,\ldots,11)$ at $c=4$
corresponds to the unifying $\w$-algebra for the second unitary
minimal models of ${\cal WA}_{n-1}$ (see below and $\q{\technical}$).
\mn
Also for ${\cal WB}_{\infty}$ we find one additional
truncation to a $\w(2,4)$ at $c=1$ which plays the same r\^ole as the
truncation of algebras of type $\w(2,3,4,\ldots)$ to $\w(2,3)$
at $c=-2$: The truncation to a $\w(2,4)$ or $\w(2,4,n)$ is a general
feature for algebras of type $\w(2,4,\ldots)$. Note that because
of the presence of the algebras ${\cal WD}_{-n}$ eq.\ (\winf.1)
also contains the truncations
${\cal WC}_{\infty} \trunc \w(2,4,\ldots,2 n (n+2))$ at $c=-n$.
Furthermore, we find the unexpected truncation
${\cal WC}_{\infty} \trunc \w(2,4,\ldots,14)$ at $c=-{1 \over 2}$.
\mn
In the case of $\w_{1+\infty}$ we have investigated null fields
up to dimension 7. We find that $\w_{1+\infty} \trunc \w(1,\ldots,n)$
at $c=n$, i.e.\ for the $n$th unitary minimal model $\q{\kac}$,
and $\w_{1+\infty} \trunc \w(1,2,3)$ at $c=-1$.
This result is consistent with the determinant formulae of
$\w_{1+\infty}$ presented recently $\q{\matsuo}$.
It is interesting to note that truncations take
place for all unitary quasi-finite representations $\q{\kac}$. Thus, the
existence of non-trivial unitary quasi-finite modules seems to imply the
truncation of $\w_{1+\infty}$ to a finitely generated algebra.
\bn
\section{\kd.\ Truncations of Casimir $\w$-algebras and the Kac determinant}
\mn
In the two previous sections we have obtained indications
for truncations of Casimir $\w$-algebras. We will now provide
further support for these truncations by inspection of the
Kac determinant.
\sn
Let ${\cal L}_k$ be a simple Lie algebra of rank $k$ over $\BC$,
$\Delta$ the set of its roots, $\rho$ ($\rho^\vee$) the sum
of its (dual) fundamental weights, $h$ ($h^\vee$) its (dual) Coxeter
number and ${\cal WL}_k$ the corresponding Casimir $\w$-algebra.
\sn
Then the Kac determinant of the vacuum Verma module ${\cal M}_N$ at
level $N$ related to ${\cal WL}_k$ is given by $\q{\bouschou}$:
$$ {\rm det} {\cal M}_N \sim
   \prod_{\beta \in \Delta}
   \prod_{m n \le N \atop m,n \in \BN}
   \left( ( \alpha_+ \rho + \alpha_- \rho^\vee, \beta)
         +({\textstyle {1 \over 2}} (\beta,\beta) m \alpha_+ + n\alpha_-)
   \right)^{p_k(N-mn)}
\eqno{\rm(\kd.1)}$$
where $p_k(x)$ is the number of partitions of $x$ into $k$ colours and
the $\alpha_{\pm}$ are related to the central charge by
$$ \alpha_+ \alpha_- = -1, \qquad
    c = k - 12 (\alpha_+\rho + \alpha_-\rho^\vee)^2.
\eqno{\rm(\kd.2)}$$
At least one $\w$-algebra singular vector has to exist
if a Casimir $\w$-algebra truncates at a certain value of
the central charge. Therefore, the Kac determinant of the
vacuum Verma module has to vanish at the level corresponding
to the singular vector.
\sn
Let us now concentrate on degenerate values of the central charge of
the Casimir $\w$-algebra where one has $\alpha_+ = {q \over \sqrt{pq}}$
and $\alpha_- = - {p \over \sqrt{pq}}$ with $p,q \in \BN$ such that
$$c_{{\cal L}_k}(p,q) = k -
12{(q\,\rho - p\,{\rho^\vee})^2 \over pq}.
\eqno{\rm(\kd.3)}$$
The condition that a singular vector occurs at level $N$
implies that one of the non-embedded factors of the Kac
determinant with $N = mn$ ($m$, $n$ positive)
is zero for a certain root $\beta$:
$$(\alpha_+ \rho + \alpha_- \rho^\vee, \beta)
 +\bigl({\textstyle{1 \over 2}}
(\beta,\beta) m \alpha_+ + n\alpha_-\bigr) = 0.
\eqno{\rm(\kd.4)}$$
Using the expressions for $\alpha_{\pm}$ this implies
$$ {p+q \over p-q} =
   {{1 \over 2} (\beta,\beta) m + n +
         (\rho,\beta) + (\rho^\vee,\beta)
   \over {1 \over 2} (\beta,\beta) m - n +
         (\rho,\beta) - (\rho^\vee,\beta)}.
\eqno{\rm(\kd.5)}$$
For given $p$ and $q$ the root leading to the lowest
singular vector is the highest root of length 2 for
all simple Lie algebras besides ${\cal C}_k$ where one has to take
the sum of the simple roots
(a table of these roots can be found in $\q{\humph}$).
Choosing  this root for $\beta$ and parametrizing
$p$ and $q$ as $p= h^\vee - 1 + r$, $q = h - 1 + s$ we obtain
$${r+s+ h+h^\vee -2 \over r-s - h+h^\vee} =
  {m  + n + h + h^\vee - 2
   \over m - n - h + h^\vee}.
\eqno{\rm(\kd.6)}$$
Obviously, $m=r$ and $n=s$ are solutions of this equation as
long as
$$r+h^{\vee} \not= s+h .
\eqno{\rm(\kd.7)}$$
If $r,s > 0$ this implies the
existence of a $\w$-algebra singular vector at level $N=r s$
for central charge $c_{{\cal L}_k}(h^\vee-1+r,h-1 + s)$. Note that
for a given value of the central charge one has to choose
minimal $p$, $q$ such that $p \ge h^{\vee}$, $q \ge h$. This
choice of not necessarily coprime integers ensures  $r,s > 0$.
\sn
A truncation takes place if the singular vector at level $N=rs$
corresponds to a non-composite field and if with the vanishing
of the simple field of dimension $N$ also all other simple fields
with dimension greater than $N$ become null fields. A singular vector
at $N=rs$ with $N$ less than the maximal spin of the generating
simple fields indicates a truncation of the Casimir algebra.
We will now discuss examples where truncations can be obtained from
this argument.
\mn
The Casimir algebra of type ${\cal WA}_n$ truncates for
$c = c_{{\cal A}_n}(n+r,n+s)$ to an algebra of type
$\w(2,3,\dots,rs-1)$. This supports the identification
(\tru.4) which corresponds to $r=1$. Furthermore, for the
$k$th unitary minimal model of ${\cal WA}_n$ this predicts
a truncation to a $\w(2,3,\ldots,k^2+3k+1)$ ($r=k+1$, $s=k+2$)
which will be established in $\q{\technical}$.
{}From the free field realization $\q{\fateev}$ it follows that
all non-composite fields of dimension greater than $N$ vanish
if the non-composite field of dimension $N$ is a null field.
Thus, for ${\cal WA}_n$ the only assumption which is not
proven yet is that the $\w$-algebra singular vector at level
$N=r s$ corresponds to a non-composite field.
\mn
The Casimir algebras related to the simple Lie algebras
of type ${\cal B}_n$ and ${\cal C}_n$ truncate for $r\Cdot s$ even
to $\w$-algebras of type $\w(2,4,\dots,rs-2)$
for $c = c_{{\cal B}_n}(2 n - 2 +r,2 n - 1+s)$
and $c = c_{{\cal C}_n}(n+r,2 n -1+s)$, respectively.
This verifies eqs.\ (\tru.8)-(\tru.15). Furthermore, it implies
the truncation of ${\cal WC}_n$ at $c_{{\cal C}_n}(m+n+1,
2 m + 2 n + 1)$ to an algebra of type $\w(2,4,\ldots,2 m (m + 2))$.
These unifying objects are in fact the algebras ${\cal WD}_{-m}$
as will be shown in $\q{\technical}$
\mn
The $\BZ_2$ orbifold of the ${\cal WD}_n$ Casimir algebra $\q{\technical}$
truncates for $c = c_{{\cal D}_n}(2 n - 3+r,2 n -3+s)$ for $r \Cdot s$ even
to an algebra of type $\w(2,4,\dots,rs-2)$. For the $k$th unitary minimal
model of $\orb{{\cal WD}_n}$
($s=r+1=k+2$) this predicts a unifying $\w$-algebra of
type $\w(2,4,\ldots,k^2+3 k)$.
This identification can be verified applying a character argument
to level-rank-duality $\q{\technical}$. The case $k=2$ is
the second line of (\tru.18).
\mn
The truncations of Casimir $\w$-algebras indicated by the Kac determinant
are summarized in table 1.
\mn
\centerline{\vbox{
\hbox{
\vrule \hskip 1pt
\vbox{ \offinterlineskip
\def\tablespace{ height2pt&\omit&&\omit&&\omit&\cr }
\def\tablerule{ \tablespace\noalign{\hrule}\tablespace }
\hrule\halign{&\vrule#&\strut\quad\hfil#\hfil\quad\cr
\tablespace
\tablespace
& {\it Casimir algebra}  && $c$ && {\it truncated algebra} &\cr
\tablerule
& ${\cal WA}_n$  && $c_{{\cal A}_n}(n+r,n+s)$
&& $\w(2,\dots,rs-1)$ &\cr
\tablerule
&${\cal WB}_n$   && $c_{{\cal B}_n}(2 n -2+r,2 n -1+s) $
&& $\w(2,4,\dots,rs-2)$ for $r\Cdot s$ even   &\cr
\tablerule
&${\cal WC}_n$   && $c_{{\cal C}_n}(n+r,2 n -1+s) $
&& $\w(2,4,\dots,rs-2)$  for $r\Cdot s$ even  &\cr
\tablerule
& $\orb{{\cal WD}_n}$   && $c_{{\cal D}_n}(2 n - 3+r,2 n - 3+s) $
&& $\w(2,4,\dots,rs-2)$   for $r\Cdot s$ even &\cr
\tablespace}
\hrule}\hskip 1pt \vrule}
\hbox{\quad Table 1: Truncations of Casimir $\w$-algebras}}
}
\bn
\section{4.\ Conclusion and outlook}
\mn
In this letter we have shown the existence of many identifications between
$\w$-algebras using a particular parametrization of the structure constants.
The identifications are closely related to truncations of $\w$-algebras.
In particular, we predicted various truncations of Casimir $\w$-algebras and
the linear $\w_\infty$-algebras.
We confirmed and generalized these truncations by inspection of the
Kac determinant.
\sn
Bits and pieces of this picture have been scattered over the literature
including a large number of misleading statements which we did not refer to.
\sn
A unifying $\w$-algebra exists at generic $c$ for each truncation in table 1
and fixed positive integers $r$, $s$ satisfying (\kd.7). This can
be established e.g.\ by regarding the structure constants of these
algebras which have fixed field content as continuous functions of
the level $n$ using the parametrizations (\tru.3) and (\tru.7).
Another argument is that for a given set of generating fields the
Jacobi identities are satisfied for all $c$ because of their polynomial
character once they are satisfied for infinitely many values of the
central charge $c$. Note that each Casimir $\w$-algebra truncates only
for finitely many values of the central charge $c$. Nevertheless one
obtains infinitely many unifying $\w$-algebras (for each $r$, $s$)
which start at a rank that increases as quickly as $r\Cdot s$ does.
\sn
Some of these unifying $\w$-algebras can be realized in terms of
cosets generalizing the notion of level-rank-duality $\q{\bogo,\altschuler}$.
Note that a coset realization automatically ensures the existence of
the $\w$-algebra for generic $c$.
We already presented a few of these realizations in this letter, further
ones will be established in $\q{\technical}$. These realizations are
summarized in table 2.
\mn
\centerline{\vbox{
\hbox{
\vrule \hskip 1pt
\vbox{ \offinterlineskip
\def\tablespace{ height2pt&\omit&&\omit&&\omit&&\omit&&\omit&\cr }
\def\tablerule{ \tablespace\noalign{\hrule}\tablespace }
\hrule\halign{&\vrule#&\strut\hskip1mm\hfil#\hfil\hskip1mm\cr
\tablespace
\tablespace
& {\it Casimir}  && {\it central charge} && {\it coset realization} &&
  {\it dimensions of} && {\it dimension of} & \cr
\tablespace
& {\it algebra}  && $c$ && {\it of unifying algebra} &&
  {\it simple fields} && {\it first null field} & \cr
\tablerule
& ${\cal WA}_{n-1}$  && $c_{{\cal A}_{n-1}}(n \pl k,n \pl k \pl 1)$ &&
 ${\widehat{su(k+1)}_n \over \widehat{su(k)}_n \oplus \widehat{U(1)}}$ &&
    $2,3,\ldots,k^2+3 k + 1$ && $k^2+3k+4$ & \cr
\tablerule
& ${\cal WA}_{n-1}$  && $c_{{\cal A}_{n-1}}(n \pl 1,n \pl 3)$ &&
 ${\w_3^{(2)} \over \widehat{U(1)}}$ &&
    $2,3,4,5,6,7$ && $10$ & \cr
\tablerule
& $\orb{{\cal WD}_n}$ && $c_{{\cal D}_n}(n \pl k \pl 1,n \pl k \pl 2)$ &&
   $\orb{ {\widehat{so(k+1)}_{2 n} \over \widehat{so(k)}_{2n} } }$ &&
    $2,4,\ldots,k^2+3 k$ && $k^2+3k+4$ & \cr
\tablerule
&${\cal WB}_n$   && ${\hbox{$c_{{\cal B}_n}(2n \pl k \hbox{$-$} 1, 2n\pl 1)$}
                     \atop
                     \hbox{$c_{{\cal B}_n}(2n , 2n \pl k)$}}$ &&
 $\hbox{(Orb)}\left({\widehat{so(k)}_\mu \oplus \widehat{so(k)}_1 \over
    \widehat{so(k)}_{\mu+1}}\right)$ &&
    $2,4,\ldots,2 k$ && $2k + 4$ & \cr
\tablerule
&${\cal WC}_n$   && $c_{{\cal C}_n}(n \pl k \pl 1,2 n \pl 2 k \pl 1)$ &&
 ${\widehat{sp(2k)}_n \oplus \widehat{sp(2k)}_{-{1 \over 2}} \over
    \widehat{sp(2k)}_{n-{1 \over 2}} }$ &&
    $2,4,\ldots,2 k^2 + 4  k$ && $2 k^2 + 4 k +5$ & \cr
\tablespace}
\hrule}\hskip 1pt \vrule}
\hbox{\quad Table 2: Coset realization of unifying $\w$-algebras}}
}
\mn
The identifications between different $\w$-algebras occur not only for
rational but also for degenerate models. There are indications
$\q{\technical}$ that all minimal models of the unifying $\w$-algebras
arise from identifications with minimal models of Casimir $\w$-algebras.
Thus, these new unifying $\w$-algebras do probably not give rise to new
minimal models and may therefore be irrelevant for the classification of
rational conformal field theories.
\sn
Looking for rational models,
it is intriguing that at least for $\w_{1+\infty}$ the existence of
non-trivial unitary quasi-finite modules seems to imply the truncation to a
finitely generated algebra.
\sn
Still, these unifying structures provide us with new insights into conformal
field theory. For example, this gives a unified approach to the conformally
invariant second order phase transition of $\BZ_n$ spin quantum chains.
They all share the same $\w(2,3,4,5)$ symmetry algebra and the growth of
the number of states with energy is always bounded by the number of partitions
into two colours. Finally, we would like to mention that these unifying
$\w$-algebras have supersymmetric generalizations (see $\q{\wisski}$ for
examples in the case $N=2$).
\bn
\section{Acknowledgments}
\mn
We are indebted to L.\ Feh\'er and W.\ Nahm for many
useful discussions and careful reading of the manuscript.
Furthermore, we are grateful to the research group of W.\ Nahm
and H.G.\ Kausch for comments during this work.
\sn
W.E.\ thanks the Max-Planck-Institut f\"ur Mathematik
in Bonn-Beuel for financial support. K.H.\ is grateful to
the University of Torino, Departement of Theoretical Physics,
for kind hospitality. R.H.\ is supported by
the NRW-Graduierten\-f\"orderung.
\vfill
\eject
\section{References}
\mn
\bibitem{\bouschou} P.~Bouwknegt, K.~Schoutens,
             {\it $\w$-Symmetry in Conformal Field Theory},
             Phys.~Rep.\ {\bf 223} (1993) p.~183
\bibitem{\kauwatts} H.G.\ Kausch, G.M.T.\ Watts,
             {\it A Study of $\w$-Algebras Using Jacobi Identities},
             Nucl.\ Phys.\ {\bf B354} (1991) p.\ 740
\bibitem{\blm} R.\ Blumenhagen, M.\ Flohr, A.\ Kliem,
             W.\ Nahm, A.\ Recknagel, R.\ Varnhagen,
             {\it $\w$-Algebras with Two and Three Generators},
             Nucl.\ Phys.\ {\bf B361} (1991) p.\ 255
\bibitem{\rva} R.\ Varnhagen, {\it Characters and Representations of
             New Fermionic $\w$-Algebras},
             Phys.\ Lett.\ {\bf B275} (1992) p.\ 87
\bibitem{\wirrep} W.\ Eholzer, M.\ Flohr, A.\ Honecker,
             R.\ H{\"u}bel, W.\ Nahm, R.\ Varnhagen,
             {\it Representations of $\w$-Algebras with Two Generators
             and New Rational Models},
             Nucl.\ Phys.\ {\bf B383} (1992) p.\ 249
\bibitem{\howcl} W.~Eholzer, A.~Honecker, R.~H\"ubel,
             {\it How Complete is the Classification
             of $\w$-Sym\-me\-tries ?}, Phys.~Lett.~{\bf B308}
             (1993) p.~42
\bibitem{\kausch} H.G.\ Kausch,
             {\it Extended Conformal Algebras Generated by a Multiplet
             of Primary Fields},
             Phys.\ Lett.\ {\bf B259} (1991) p.\ 448
\bibitem{\flohr} M.~Flohr, {\it $\w$-Algebras,
             New Rational Models and Completeness of the $c=1$
             Classification}, Commun.~Math.~Phys.~{\bf 157}
             (1993) p.~179
\bibitem{\bowwatts} P.\ Bowcock, G.M.T.\ Watts,
             {\it On the Classification of Quantum $\w$-Algebras},
             Nucl.\ Phys.\ {\bf B379} (1992) p.\ 63
\bibitem{\FORT} L.\ Feh\'er, L.\ O'Raifeartaigh, P.\ Ruelle, I.\ Tsutsui,
             {\it On the Completeness of the Set of Classical $\w$-Algebras
             Obtained from DS Reductions}, preprint BONN-HE-93-14 (1993),
             DIAS-STP-93-02, hep-th/9304125,
             to appear in Commun.\ Math.\ Phys.\
\bibitem{\fehort} L.\ Feh\'{e}r, L.\ O'Raifeartaigh, I.\ Tsutsui,
             {\it The Vacuum Preserving Lie Algebra of a Classical
             $\w$-Algebra}, Phys.\ Lett.\ {\bf B316} (1993) p.\ 275
\bibitem{\dBT} J.\ de Boer, T.\ Tjin, {\it The Relation between Quantum
             $\w$ Algebras and Lie Algebras},
             Commun.\ Math.\ Phys.\ {\bf 160} (1994) p.\ 317
\bibitem{\bbss} F.A.\ Bais, P.\ Bouwknegt, M.\ Surridge, K.\ Schoutens,
             {\it Extensions of the Virasoro Algebra Constructed
             from Kac-Moody Algebras Using Higher Order Casimir Invariants};
             {\it Coset Construction for Extended Virasoro Algebras},
             Nucl.\ Phys.\ {\bf B304} (1988) p.\ 348; p.\ 371
\bibitem{\ajl} J.~de Boer, L.~Feh\'{e}r, A.~Honecker,
             {\it A Class of $\w$-Algebras with Infinitely Generated
             Classical Limit}, BONN-HE-93-49,
             ITP-SB-93-84, hep-th/9312049,
             to appear in Nucl.~Phys.~{\bf B}
\bibitem{\hornfeck} K.~Hornfeck,
             {\it $\w$-Algebras with Set of Primary
             Fields of Dimensions $(3,4,5)$ and $(3,4,5,6)$},
             Nucl.~Phys.~{\bf B407} (1993) p.~237
\bibitem{\klausREP} K.~Hornfeck, {\it Classification \Skp of \Skp
             Structure \Skp Constants \Skp for \Skp $\w$-Algebras \Skp
             from \Skp Highest \Skp Weights},
             Nucl.\ Phys.\ {\bf B411} (1994) p.~307
\bibitem{\technical} R.\ Blumenhagen, W.\ Eholzer, A.\ Honecker,
             K.\ Hornfeck, R.\ H\"ubel, {\it Coset Realization of
             Unifying $\w$-Algebras}, in preparation
\bibitem{\bogo} P.\ Bowcock, P.\ Goddard, {\it Coset Constructions
             and Extended Conformal Algebras},
             Nucl.\ Phys.\ {\bf B305} (1988) p.\ 685
\bibitem{\altschuler} D.\ Altschuler, {\it Quantum Equivalence of Coset
             Space Models}, Nucl.\ Phys.\ {\bf B313} (1989) p.\ 293
\bibitem{\poly} A.M.\ Polyakov, {\it Gauge Transformations
             and Diffeomorphisms},
             Int.\ Jour.\ of Mod.\ Phys.\ {\bf A5} (1990) p.\ 833
\bibitem{\bersh} M.\ Bershadsky, {\it Conformal Field Theories
             via Hamiltonian Reduction}, Commun.\ Math.\ Phys.\
             {\bf 139} (1991) p.~71
\bibitem{\altbausal} D.\ Altschuler, M.\ Bauer, H.\ Saleur,
             {\it Level-Rank Duality in Non-Unitary Coset Theories},
             J.~Phys.~A: Math.~Gen.~{\bf 23} (1990) p.~L789
\bibitem{\bouwknegt} P.~Bouwknegt, {\it Extended Conformal
             Algebras from Kac-Moody Algebras},
             Proceedings of the meeting `Infinite dimensional
             Lie algebras and Groups'
             CIRM, Luminy, Marseille (1988) p.~527
\bibitem{\andrdipl} A.\ Honecker, {\it Darstellungstheorie von
             $\w$-Algebren und Rationale Modelle
             in der Konformen Feldtheorie},
             Diplomarbeit BONN-IR-92-09 (1992)
\bibitem{\popeo} C.N.\ Pope, L.J.\ Romans, X.\ Shen,
             {\it $\w_\infty$ and the Racah-Wigner Algebra},
             Nucl.\ Phys.\ {\bf B339} (1990) p.\ 191
\bibitem{\popet} C.N.\ Pope, L.J.\ Romans, X.\ Shen,
             {\it A New Higher-Spin Algebra and the Lone-Star Product},
             Phys.\ Lett.\ {\bf B242} (1990) p.\ 401
\bibitem{\kac} V.\ Kac, A.\ Radul, {\it Quasifinite Highest
             Weight Modules over the Lie algebra of Differential
             Operators on the Circle},
             Commun.~Math.~Phys.~{\bf 157} (1993) p.~429
\bibitem{\matsuo} H.\ Awata, M.\ Fukuma, Y.\ Matsuo, S.\ Odake,
             {\it Determinant
             Formulae of Quasi-Finite Representations of $\w_{1+\infty}$
             Algebra at Lower Levels}, preprint YITP/K-1054, UT-669,
             SULDP-1994-1, hep-th/9402001
\bibitem{\humph} J.E.\ Humphreys, {\it Introduction to Lie Algebras and
             Representation Theory}, Springer-Verlag, New York, Heidelberg,
             Berlin (1972)
\bibitem{\fateev} V.A.~Fateev, S.L.~Lukyanov, {\it The Models of
             Two-Dimensional Conformal Quantum Field Theory
             with $\BZ_n$ Symmetry},
             Int.~Jour.~of Mod.~Phys.~{\bf A3} (1988) p.~507
\bibitem{\wisski} R.\ Blumenhagen, A.\ Wi{\ss}kirchen,
             work in progress
\vfill
\end